\documentclass[prl,twocolumn,showpacs,preprintnumbers,amsmath,amssymb]{revtex4}

\usepackage{graphicx}
\usepackage{dcolumn}
\usepackage{bm}

\begin{document}

\preprint{...}

\title{Signal buffering in random networks of spiking neurons: \\microscopic vs. macroscopic phenomena}

\author{Julien Mayor}
\author{Wulfram Gerstner}
\email{wulfram.gerstner@epfl.ch}
\affiliation{Brain-Mind Institute and School of Computer and Communication Sciences\\
Ecole Polytechnique F\'{e}d\'{e}rale de Lausanne (EPFL), CH-1015 Lausanne, Switzerland
}
\date{\today}
            
\begin{abstract}
In randomly connected networks of pulse-coupled elements
a time-dependent input signal can be buffered
over a short time. We studied the signal buffering
properties in simulated networks as a function
of the networks state, characterized by
both the Lyapunov exponent of the microscopic
dynamics and the macroscopic activity
derived from mean-field theory.
 If all network elements
receive the same signal, signal buffering over delays
comparable to the intrinsic time constant
of the network elements
can be explained by macroscopic properties and
works best
at the phase transition to chaos.
However, if only 20 percent of the network
units receive a common time-dependent signal,
signal buffering properties improve and can
no longer be attributed to the macroscopic dynamics.
\end{abstract}

\pacs{87.18.Sn, 87.10.+e, 82.40.Bj}
\maketitle
\section{Introduction}
Neurons in the brain are connected with each
other and send
short electrical pulses (action potentials or spikes)
along those connections.
Despite the fact that there are 
correlations between the type of connections
and the type of neurons \cite{Gupta00},
it is fair to say that neurons
fall essentially into two classes,
excitatory and inhibitory,
and that
the connectivity in a local population
of  several thousand
cortical neurons  is close to  random.
Networks with fixed random connectivity
can, in principle, contain
loops of varying size,
which could sustain  the flow
of transient information signals
over times that are long
compared to the intrinsic
time constants of the network
elements, i.e., the neurons.
In neuroscience and related fields,
elementary  considerations on information
flow in random networks
have inspired ideas
as diverse as synfire
chain activity \cite{Abeles91},
reverberations \cite{Kistler02c},
liquid computing \cite{MaassETAL:01a},
echo state machines \cite{Jaeger04},
or computing at the edge of chaos
\cite{Bertschinger04,Langton90}.

On the other hand, random networks
have also been studied intensively by
the physics community,
in the context of diluted spin glasses
\cite{Viana85},
formal neural networks
\cite{Amari72,Derrida87}
or automata \cite{Derrida86}
and limiting cases have been identified
for which exact solutions are known.
In particular,  in the limit
of asymmetric networks with low connectivity,
 mean-field dynamics
becomes exact \cite{Derrida87}.
More recently these approaches have been extended
to the case of random networks of spiking
neurons in continuous time
\cite{Brunel99}.

In this paper, we will compare
simulations of a random network of excitatory
and inhibitory neurons
with the mean-field solutions
valid in the low-connectivity limit
and evaluate the performance
of such networks
on a simple
information buffering task
that can be seen as
a minimal and necessary requirement
for more complex computational tasks
\cite{MaassETAL:01a,Jaeger04} which
a neural network might have to solve.
More precisely, the task consists
in reconstructing a time-dependent
 input $I(t')$ by reading out  the activity
of the network at a later time
$t=t'+\Delta$.
We will see that
performance in the information
buffering task is best
at the phase transition 
that is marked by a rapid
increase in both the
\emph{macroscopic} activity
variable and the
Lyapunov exponent characterizing
the \emph{microscopic} state
indicating transition to chaos.
Surprisingly,
if the same time-dependent input $I(t)$ is shared by
all neurons in the network, an information  readout based
only on the  macroscopic variable
is as good as a readout that
is based on the output pulses of all $N$ neurons.
However, if the input is only given
to a small group of neurons
a detailed readout conveys more information than a macroscopic one 
suggesting that loops in the network
connectivity might indeed play a role.

\section{Model}
We consider a network of $N$ leaky integrate-and-fire units
(neurons)
with fixed random connectivity.
 80 percent of the neurons are taken as excitatory
and the remaining 20 percent  inhibitory.
Independent of the network size ($N=200,400,800$),
each neuron $i$ in our simulation receives input from
$C_E=40$ excitatory and $C_I=10$ inhibitory units  (presynaptic neurons),
 which are chosen at random amongst the $N-1$ other neurons in the
network.
The ensemble of neurons that are presynaptic to neuron $i$ is denoted by $M_i$
and the efficacy $w_{ij}$ of a connection from
a presynaptic neuron $j\in M_i$ to a postsynaptic neuron
$i$ is $w_{ij}=w_E$ if $j$ is excitatory,
and $w_{ij}=w_I$ if $j$ is inhibitory.

Each neuron is described by a linear equation 
combined with a threshold.
In the subthreshold regime
the membrane potential  follows the differential equation 
\begin{equation}\label{if}
    \tau_m \dot{u_i}(t) = -u_i(t) + I_i^{\rm netw} + I_i^{\rm ext}(t)
\end{equation}
where
$\tau_m = 20$ms is the effective
membrane time constant and $I_i^{\rm ext}$ external input.
The recurrent input $I_i^{\rm netw}$  neuron $i$ receives from the network
is
\begin{equation}
    I_i(t)= \tau_m \sum_{j \in M_i} \omega_{ij} \sum_k \delta(t-t_j^k-D)
\end{equation}
where $t_j^k$ is the
time neuron $j$ fires its $k$th spike and $D=1$ms  is a short  transmission
delay. A spike from   an excitatory (inhibitory) neuron $j\in M_i$
causes a jump in the membrane potential of neuron $i$ of
 $w_E=0.6$mV ($w_I=-3.6$mV).
If the membrane potential $u_i$ reaches
a threshold $\vartheta = 10$mV,
a spike of neuron $i$ is recorded and
its membrane potential is reset to
$u_i=0$. Integration restarts after an absolute refractory period of $\tau_{\rm rp}=2$ms.

The external input $I_i^{\rm ext}$ can be separated into
two components. First, we inject a time dependent test signal
$I_i^{\rm sg}(t)$ which is generated as follows: the total 
simulation time is broken into segments of duration $T_{\rm sg}=10$ms.
During each segment of length $T_{\rm sg}$ the input is kept
constant. At the transition to the next segment, a new value of
$I^{\rm sg}(t)$ is drawn from a uniform distribution
over the interval $[-0.25,0.25]$, i.e. the signal distribution has a 
standard deviation of $\sigma_{\rm sg}=\frac{0.25}{\sqrt{3}}=0.144$.
By construction the signal at time $t$ provides no information about the
signal at $t-T$ for $T>T_{\rm sg}$. More precisely, the autocorrelation
$A(s)$ of the signal has a triangular shape and is strictly zero for $|s|>T_{\rm sg}$.

Second, the network
of $N$ neurons is considered as part of a larger
brain structure. To mimic input from excitatory `background' neurons
that are not modelled explicitly,
we assume stochastic spike arrival described
by a Poisson process of total rate $\nu_{\rm exc}$.
For the sake of simplicity,
we assume that the efficacy $w_E$ of background spikes
is identical to that of the recurrent connections
within the network.
We approximate the background input
$I^{\rm backg}$ by a Gaussian white noise with mean
 $\mu^{\rm backg}= \omega_E \nu_{\rm exc} \tau_m$ and
standard deviation
$\sigma^{\rm backg}=\omega_E \sqrt{\nu_{\rm exc}\tau_m}$, where $\nu_{\rm exc}$ is the 
background spike arrival rate. 
This approximation is valid under the assumption that a neuron receives a large number of
presynaptic contributions per unit time, each generating a change
in the membrane potential that is relatively
small compared to the firing threshold.
To simplify notation we set $\gamma=C_I/C_E$ and $g=-\omega_I/\omega_E$.
The values in our simulations are $\gamma=0.25$ and $g=6$.

\section{Characterizing the network state}

\begin{figure}[!hbt]
\includegraphics*[angle=0, scale=0.21]{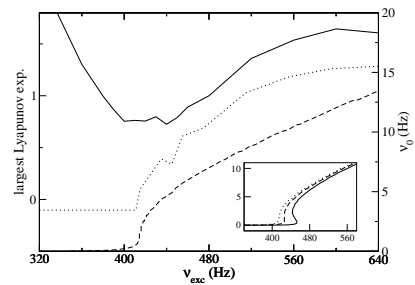}
\caption{Dashed line: Fixed points of the population activity $\nu_{0}$ as a function of the Poisson background drive 
resulting from Eqs. (\ref{input-mean}-\ref{eq2}). Note the sharp transition at $\nu_{\rm exc}= 420$Hz. 
The network switches 
from an almost quiescent state to a state of sustained activity.
Dotted line: largest Lyapunov exponent as a function of the external drive $\nu_{\rm exc}$ in a network of 800 neurons. 
For $\nu_{\rm exc}< 420$Hz the population rate $\nu_0$ is low ('quiescent state') and the largest Lyapunov exponent 
is negative. 
For a stronger drive, the exponent switches to a positive value, reflecting the chaotic
behaviour of the membrane potential trajectories. 
Solid line: Signal reconstruction error for a delay of 20 ms in a network of 800 neurons. The error is minimal near the
transition from the quiescent to a chaotic regime. Inset: Fixed points of the population rate (Eq. \ref{eq2}) in absence of test signal (solid line),
and with increasing signal variance. The system exhibit a first-order phase transition if the signal is weak.
}\label{Fig-transition}
\end{figure}

If we replace the signal $I^{\rm sg}$
by  Gaussian white noise of
zero mean and standard deviation $\sigma_{\rm sg}$
a mean-field analysis
of the random network defined above
can be performed following
\cite{Brunel99}.
The macroscopic variable describing
the activity of the network is the
population rate $\nu_0$. In a stationary
state of asynchronous neuronal activity, the population rate
depends on the mean
\begin{equation}
\label{input-mean}
    \mu_0 = \omega_E \tau_m (\nu_{\rm exc} + C_E \,\nu_0(1-g\gamma))
\end{equation}
and variance
\begin{equation}
\label{input-variance}
    \sigma^2_0 = \omega^2_E \tau_m ( \nu_{\rm exc} + C_E\,\nu_0(1+g^2\gamma)) + \sigma^2_{\rm sg}
\end{equation}
of the total input $I^{\rm netw} + I^{\rm ext}$ via the equation
\begin{equation}
\frac{1}{\nu_0}=\tau_{\rm rp}+2\pi\int_{\frac{-\mu_0}{\sigma_0}}^{\frac{\vartheta-\mu_0}{\sigma_0}}due^{u^2}\int_{-\infty}^{u}dve^{-v^2}
\label{eq2}
\end{equation}
The mean-field result is valid in the low connectivity limit \cite{Brunel99}.
Combining Eqs. (\ref{input-mean}) - (\ref{eq2}),
we obtain
a self-consistent solution of
the population firing rate $\nu_0$ as a function of the external Poisson drive $\nu_{\rm exc}$. The
population rate shows a marked increase 
near $\nu_{\rm exc} \cong 420$Hz as shown in Fig. \ref{Fig-transition}
which is in the vicinity of a first-order phase transition
predicted by the mean-field theory. If we decrease the amplitude of
the input signal $\sigma_{\rm sg} \to 0$ the phase transition becomes more pronounced and 
a regime of coexistence of several solutions appears (inset).

The marked increase in the macroscopic variable
$\nu_0$ is accompanied by a transition
of the largest Lyapunov exponent from
negative to positive values
which indicates that the microscopic dynamics
becomes chaotic.
The largest Lyapunov exponent is
defined by $\lambda_p=\lim_{t\rightarrow \infty} ln\frac{\epsilon(t)}{\epsilon(0)}$
where
$\epsilon(t)=\sum_{i=1}^N
(u_i(t)-u'_i(t))^2$
is the difference between a reference trajectory
of the $N$ variables $u_1(t),...u_N(t)$
and test trajectory $u_1'(t),...u_N(t)'$
with slightly different initial conditions.
Using standard numerical techniques
\cite{Sprott03}, the largest Lyapunov exponent
has been estimated from
a large number of simulations
of a test network and a reference network
with identical connectivity.
Both networks received identical
input (same realization of the
Poisson input and signal input)
and the test trajectory was
regularly reset close to the reference
trajectory.
Moreover, we confirmed that not only for background spike input 
but also for appropriately chosen
constant input (i.e. $\mu^{\rm backg}>0$, $\sigma^{\rm backg}=0$),
the fixed random connectivity is sufficient
to generate irregular asynchronous
spiking activity \cite{Brunel99}
with a positive
largest Lyapunov exponent (data not shown).

\section{Information buffering}

After having characterized the macroscopic
and microscopic state of the network,
we asked how
performance in an information
buffering task, inspired by
concepts of liquid computing
\cite{MaassETAL:01a}
and echo state machines \cite{Jaeger04},
would depend on the network state.
We considered a linear readout unit
with dynamics
$ dy/dt = -((y-\alpha_0)/\tau_s) + \sum_{i=1}^N \alpha_i \sum_k \delta(t-t_i^k)$
where the sums run over all firing times $t_i^k$ of
all neurons in the network. $\tau_s=5$ms is a short synaptic time constant.
The $N+1$ free parameters
$\alpha_i$ ($0\le i\le N$)  are chosen so as to minimize
the signal reconstruction error
$E = \langle [y(t) - I^{\rm sg}(t-\Delta)]^2\rangle /\sigma_{\rm sg}^2$.
Parameters were optimized using
a first simulation (learning set)
 lasting 100 seconds (100'000 time steps of simulation)
and were kept fixed afterwards.
The performance measurements reported in this paper are then
evaluated on a second simulation of 100 seconds (test set). 
Simulation results were obtained using the simulation software 
NEST\footnote{NEST Initiative, available at www.nest-initiative.org}.

\begin{figure}[!hbt]
\includegraphics*[angle=0,scale=1]{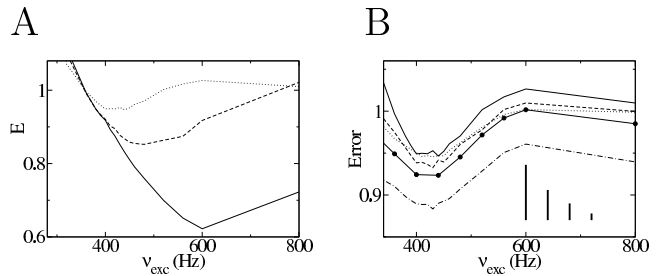}
\caption{A. Signal reconstruction error $E$ as a function of the background firing rate 
in a network of 800 neurons for three different information buffer delays: 
$\Delta=10$ms (solid), $\Delta=15$ms (dashed), $\Delta=20$ms (dotted). 
For sufficiently long delays, optimal performance is located near the transition
between the quiescent and the chaotic state; cf Fig \ref{Fig-transition}. 
Deeper in the chaotic phase the error goes back to the chance level whereas in the almost quiescent regime we can see 
the effects of overfitting ($E>1$), because the number of action potentials is insufficient to build 
an accurate model of the past events.
B. Comparison of the errors for different network sizes: $N=200,400,800$ neurons (resp. dotted, dashed and solid lines), 
for a delay $\Delta=20$ms. The solid line with filled circles corresponds to the \emph{macroscopic} readout of the network of 
$N=800$ neurons.
The location of minimal error is independent of the number of neurons and coincides with the phase transition.
The vertical shift of the error curves is not significant but due to overfitting because of limited amount of data.
Vertical bars indicate the mean difference between errors on the data used for parameter optimisation (training set) 
and that on an independent test set for $N=800$ (left bar), $N=400$ (2nd bar), $N=200$ (3rd bar) and macroscopic 
readout (right bar). A representative curve of errors on the training set for $N=800$ neurons is shown by the dot-dashed line.
}
\label{Fig-delays}
\end{figure}

The same time-dependent signal $I^{\rm sg}(t)$
was injected into all neurons in the network
and the performance evaluated in terms
of the signal reconstruction error.
The performance
depends on the delay $\Delta$ of information buffering
which has to be compared with the membrane time
constant ($\tau_m=20$ms) and the
autocorrelation of the input $T_{\rm sg} = 10$ms. Overall the signal reconstruction 
error is relatively high.
As expected, the signal reconstruction
error increases if we increase the desired buffer duration from
$\Delta=10$ms to $\Delta=15$ms or $\Delta=20$ms
(Fig. \ref{Fig-delays}A).
At the same time, the optimal background firing
rate $\nu_{\rm exc}$ to achieve minimal
signal reconstruction error shifts towards
lower values and is for
$\Delta=20$ms very close to the
transition between regular and chaotic microscopic dynamics
as shown in Fig. \ref{Fig-transition}.
This result is consistent with the idea
of computing at the edge of chaos
in cellular automata \cite{Langton90}
or threshold gates \cite{Bertschinger04}.
Also, similar to the results in
discrete-time spin networks \cite{White04},
the information buffering performance does not significantly
depend on the number $N$ of neurons in the network;
cf. Fig. \ref{Fig-delays}B. Differences are within
the statistical variations caused by  overfitting
on finite data samples.

\section{Microscopic vs. Macroscopic properties}

Given that networks states have been
classified successfully by  macroscopic mean-field
equations \cite{Brunel99}, we wondered
whether the performance in the above
information buffering task can be completely
understood in terms of \emph{macroscopic}
quantities.
To answer this question, we compared
the performance using
the previous readout unit $y$
(i.e., a `microscopic' readout with  $N+1$ free parameters,
one per neuron plus an offset) with that 
by a simplified readout
$\tilde{y}$ with two free
parameters $\tilde{\alpha}$ and $\tilde{\beta}$ only,
$ d\tilde{y}/dt = -((\tilde{y}-\tilde{\beta})/\tau_s) +
\tilde{\alpha} \sum_{i=1}^N  \sum_k \delta(t-t_i^k)$,
i.e., a readout that uses only
the macroscopic population activity.
Surprisingly, for a stimulation paradigm
where all neurons receive the same time-dependent
signal $I^{\rm sg}(t)$ the macroscopic readout
performs as well as the microscopic one.
In other words,
connectivity loops between \emph{specific}
subsets of neurons where signals
could circulate for some time
seem not to play a role
in information buffering.
This suggests that, for signals of sufficiently
small amplitude, the information buffering
capacity is directly related to
the \emph{macroscopic} linear response kernel
$\kappa$ of the network activity,
that can, in principle,
be calculated from the linearized
mean-field equations of the population rate,
i.e., $\Delta \nu(t) = \int \kappa(s) \, I^{\rm sg}(t-s) ds$;
cf. \cite{Brunel99}.
The time constant of the kernel, and hence information
buffering delays, become large in
the vicinity of a phase transition.

\begin{figure}[!hbt]
  \includegraphics*[angle=0, scale=0.21]{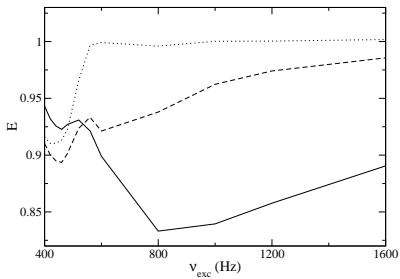}
  \caption{The input is injected to $20\%$ of the neurons only. A macroscopic 
  readout assuming a single population (dotted line) performs well near the phase transition. 
  However deeper in the chaotic phase it is outperformed by the \emph{microscopic} 
  readout (solid line). 
  A macroscopic readout based on a two-population assumption (dashed line) explains 
  only part of the increased performance. The signal buffering delay for this figure  
  is $\Delta=20$ms.
  }
  \label{fig-twogroups}
\end{figure}

We hypothesized that signal transmission
loops in our randomly connected network
could manifest themselves more easily
if only a small subset of neurons
received the input signal.
We therefore selected 20 percent
of neurons at random
(group $G_1$) and
injected an identical
time dependent signal
$I_j^{\rm sg}(t)$
into all neurons $j\in G_1$.
The remaining 80 percent of neurons
(group $G_2$) received no signal.
In such a network consisting of two groups,
signal buffering performance
is indeed significantly better
than in a single homogeneous group (Fig. \ref{fig-twogroups}).

On a macroscopic scale, a network
consisting of two groups $G_1$
and $G_2$ can be described
by two macroscopic variables,
i.e., the population activities
in groups $G_1$ and $G_2$.
In order to evaluate the information
contained in the macroscopic
population rates,
we used a linear readout unit
$y_2$ with three
free parameters $\beta_{0}$, $\beta_{1}$
and $\beta_2$, characterized by
the differential equation
$ dy_2/dt = -((y_2-\beta_{0})/\tau_s) +
\beta_1 \, \sum_{i\in G_1}  \sum_k \delta(t-t_i^k)
+
\beta_2 \, \sum_{i\in G_2}  \sum_k \delta(t-t_i^k)
$ and proceeded as before.
As we can see from
Fig. \ref{fig-twogroups},
a readout based on
the macroscopic activity of the two groups
performs significantly worse
than the microscopic readout.
This suggests, that for the case
when only a small subset of units in a random
network receive an input,
signal transmission loops,
and hence microscopic neuronal dynamics,
indeed play a role in
short-term information buffering.
\section{Discussion}
Mean-field methods neglect correlations
in the input. In random networks
mean-field theory becomes
asymptotically correct
only in the low-connectivity limit
where the probability of closed
loops tends to zero
\cite{Derrida87}.
However, it is exactly these loops
which could give the network 
the power to buffer information for
times significantly longer than
the intrinsic time constants of the network
elements.
Our network is formally
not in the low-connectivity limit
since the number of neurons
$N=800$ is small.
Nevertheless, we found that
mean-field results
can qualitatively predict
the rough location of
the
phase transition
of the macroscopic population rate.
Moreover,
if the input signal is shared between all neurons,
a macroscopic readout is sufficient to explain
the network performance in an information buffering task.
Microscopic properties do,
 however, play a role if the input is only given
to a subset of the network units suggesting
that in this case ultra-short term information
buffering in connectivity loops is indeed possible.
In additional simulations we checked that the maximum
delay $\Delta$ for which signal reconstruction is feasible
is only in the range of 20-50ms and hence not
significantly different from the intrinsic neuronal
time constants. This suggests that, without
slow processes such as synaptic plasticity or
neuronal adaptation a purely random network of spiking neurons
is not suitable as an information buffer beyond
tens of milliseconds.

\end{document}